\documentclass[a4paper,12pt]{article}

\usepackage{zref-abspage}

\usepackage[utf8]{inputenc} % Kenneth: Original document specified utf8x encoding (\usepackage[utf8x]{inputenc}) but this may cause problems without that particular encoding; use utf8 to be safe.
\usepackage[T1]{fontenc}
\usepackage{times}
\usepackage[colorlinks=true,linkcolor=black,citecolor=black,urlcolor=blue]{hyperref}
\usepackage{geometry}
\usepackage{titlesec}
\titleformat{\section}
{\bfseries\uppercase}{\thesection.}{1em}{}
\titleformat{\subsection}
{\bfseries}{\thesection.\thesubsection.}{1em}{}

\usepackage{graphicx} % used to insert the figure
\usepackage{multirow} % used for the table
\usepackage{cite}
\usepackage{breakurl}
\usepackage{indentfirst}
\usepackage{amsmath, amssymb, amsfonts, bm}
\usepackage{txfonts}
\usepackage[inline]{enumitem}
\usepackage{xcolor}
\usepackage{enumitem}

\usepackage{booktabs}
\usepackage{tabularx}
\usepackage[flushmargin, bottom]{footmisc}
\usepackage{bigfoot}
\DeclareNewFootnote[para]{default}
\DeclareNewFootnote[para]{B}[fnsymbol]
\MakePerPage{footnote}
\usepackage{mdframed}
\titlespacing*{\subsection}{0pt}{1.5em}{0.2em}
\titlespacing*{\section}{0pt}{1.5em}{0.2em}

\renewcommand\eqref[1]{Equation~\ref{#1}}

\renewcommand{\thesection}{\arabic{section}}
\renewcommand{\thesubsection}{\arabic{subsection}}

\makeatletter
\renewcommand\@biblabel[1]{#1.}
\makeatother

\newlength{\bibitemsep}
\newlength{\bibparskip}
\let\oldthebibliography\thebibliography
\renewcommand\thebibliography[1]{%
  \oldthebibliography{#1}%
  \setlength{\parskip}{\bibitemsep}%
  \setlength{\itemsep}{\bibparskip}%
}

\usepackage{siunitx}
\usepackage{cleveref}
\usepackage{caption}
\usepackage{subcaption}

\DeclareSIUnit \decibelA {dB(A)}

\makeatletter
\newcommand{\defcand}[2]{%
    \@namedef{tt#1}{\texttt{#2}}%
}
\makeatother

\defcand{amb}{ambience}
\defcand{best}{best}
\defcand{bprior}{bprior}
\defcand{wprior}{wprior}
\defcand{amss}{amss}
\defcand{amssme}{amssme}
\defcand{rand}{random}

\begin{document}

% Minimum 4 pages, maximum 12 pages

\begin{center}
	\includegraphics[width=3.50in]{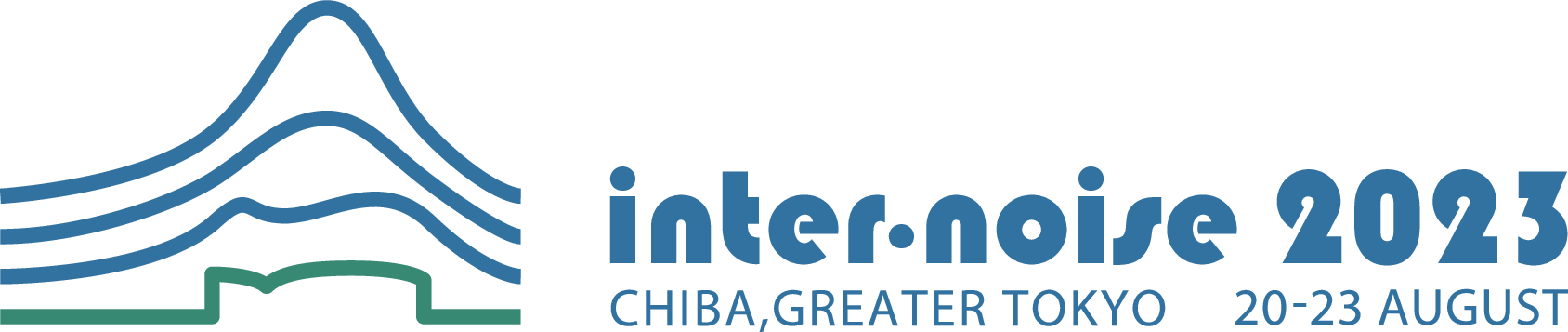}
\end{center}
\vskip.5cm

\begin{flushleft}
\fontsize{16}{20}\selectfont\bfseries
%\textcolor{red}{(The title should be written in "Times New Roman", 16-point, bold font. The first letter of the first word in the title should be capitalized)} \\
Preliminary investigation of the short-term in situ performance of an automatic masker selection system
\end{flushleft}
\vskip1cm

\renewcommand\baselinestretch{1}
\begin{flushleft}

Bhan Lam\footnote{bhanlam@ntu.edu.sg},
Zhen-Ting Ong\footnote{ztong@ntu.edu.sg},
Kenneth Ooi\footnote{wooi@e.ntu.edu.sg},
Wen-Hui Ong \footnote{wong135@e.ntu.edu.sg},
Trevor Wong \footnote{trev0006@e.ntu.edu.sg}, % Staff account (now defunct) was trevor.wong@ntu.edu.sg
Woon-Seng Gan \footnote{ewsgan@ntu.edu.sg}\\
School of Electrical and Electronic Engineering, Nanyang Technological University, Singapore\\
50 Nanyang Avenue, Singapore 639798\\

\vskip.5cm
Karn N. Watcharasupat\footnote{kwatcharasupat@gatech.edu}\\
Center for Music Technology, Georgia Institute of Technology, USA\\
840 McMillan Street NW, Atlanta, GA 30332\\

\end{flushleft}
\textbf{\centerline{ABSTRACT}}\\
\textit{Soundscape augmentation or ``masking'' introduces wanted sounds into the acoustic environment to improve acoustic comfort. Usually, the masker selection and playback strategies are either arbitrary or based on simple rules (e.g. –3 dBA), which may lead to sub-optimal increment or even reduction in acoustic comfort for dynamic acoustic environments. To reduce ambiguity in the selection of maskers, an automatic masker selection system (AMSS) was recently developed. The AMSS uses a deep-learning model trained on a large-scale dataset of subjective responses to maximize the derived ISO pleasantness (ISO 12913-2). Hence, this study investigates the short-term in situ performance of the AMSS implemented in a gazebo in an urban park. Firstly, the predicted ISO pleasantness from the AMSS is evaluated in comparison to the in situ subjective evaluation scores. Secondly, the effect of various masker selection schemes on the perceived affective quality and appropriateness would be evaluated. In total, each participant evaluated 6 conditions: (1) ambient environment with no maskers; (2) AMSS; (3) bird and (4) water masker from prior art; (5) random selection from same pool of maskers used to train the AMSS; and (6) selection of best-performing maskers based on the analysis of the dataset used to train the AMSS.} 

\section{INTRODUCTION}
\noindent
Improving acoustic comfort through soundscape augmentation or "masking" has gained significant attention in the field of noise control \cite{Brown2010TheDesign,Hao2016,Galbrun2014,Hong2020b,Hong2021b,Hong2020h}. Moreover, mounting evidence suggests that incorporating natural sound or biophillic maskers to existing soundscapes can positively impact affect and health outcomes \cite{Buxton2021, VanRenterghem2020, Coensel2011, Leung2017, Jeon2010}. However, the selection and playback strategies of maskers are often arbitrary or based on simple rules, potentially leading to sub-optimal enhancements or even discomfort in dynamic acoustic environments \cite{Ooi2023a}. To address this issue and reduce ambiguity in masker selection, an automatic masker selection system (AMSS) has been recently developed \cite{Ooi2022,Watcharasupat2022,Ooi2023b}. The AMSS utilizes a deep-learning model trained on a large-scale dataset (ARAUS \cite{Ooi2023a}) of subjective responses to maximize the derived ISO pleasantness  \cite{InternationalOrganizationforStandardization2019}.

In this study, we focus on evaluating the short-term in-situ performance of the AMSS implemented in a gazebo within an urban park. Firstly, we compare the predicted ISO pleasantness scores from the AMSS with in-situ subjective evaluation scores. By assessing this alignment, we gain insights into the effectiveness of the AMSS in capturing the perceptual qualities of the soundscape. Secondly, we investigate the impact of various masker selection schemes on the perceived affective quality and appropriateness. We evaluate six conditions for each participant: (1) ambient environment with no maskers, (2) AMSS-selected maskers \cite{Wong2022}, (3) bird masker from prior art \cite{Hong2021b}, (4) water masker from prior art \cite{Hong2021b}, (5) random selection from the same pool of maskers used to train the AMSS \cite{Ooi2023a}, and (6) selection of the best-performing masker from the ARAUS dataset \cite{Ooi2023a}.

This study aims to shed light on the performance and suitability of the AMSS in real-world settings. By comparing the predicted ISO pleasantness scores with subjective evaluations, we examine the system's ability to accurately capture the perceived pleasantness of the soundscape and its recommended masker in further enhancing the ISO pleasantness. Additionally, we explore the influence of different masker selection schemes on affective quality and appropriateness. As this is an in-situ evaluation in a dynamic sound environment, a repeated measures design was adopted.

Specifically, we investigated (1) the difference in ISO pleasantness between the AMSS predictions and in-situ participant evaluations, and (2) the effect of the dynamic in-situ ambient environment in the gazebo and masker selection schemes on ISO pleasantness, ISO eventfulness, and appropriateness.

\section{METHOD}

\subsection{Site Selection and Participants}
\noindent

\begin{figure}{}
     \centering
     \begin{subfigure}[b]{0.45\textwidth}
         \centering
         \includegraphics[width=\textwidth]{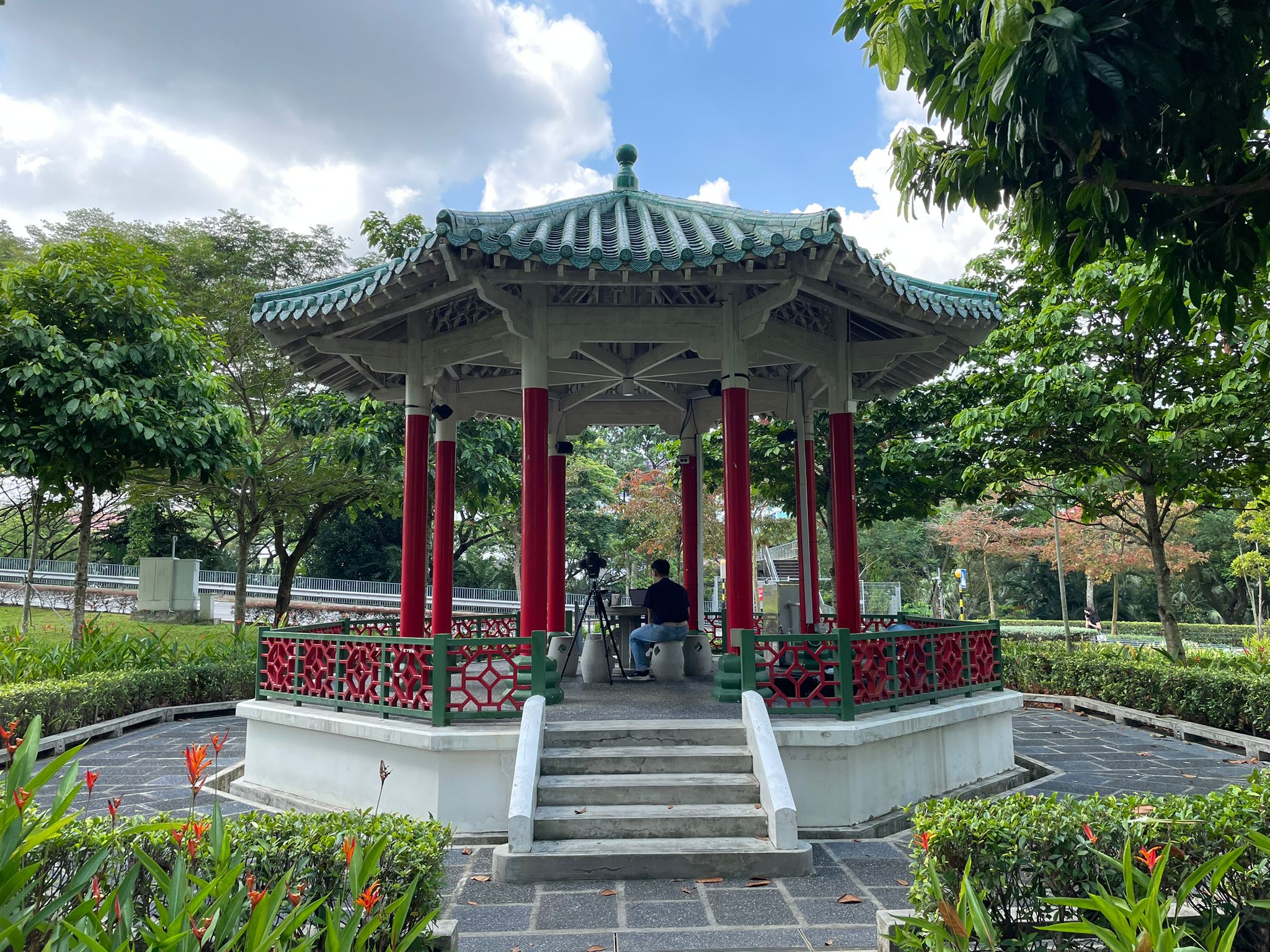}
         \caption{}
         \label{fig:sitefar}
     \end{subfigure}
     \hfill
     \begin{subfigure}[b]{0.45\textwidth}
         \centering
         \includegraphics[width=\textwidth]{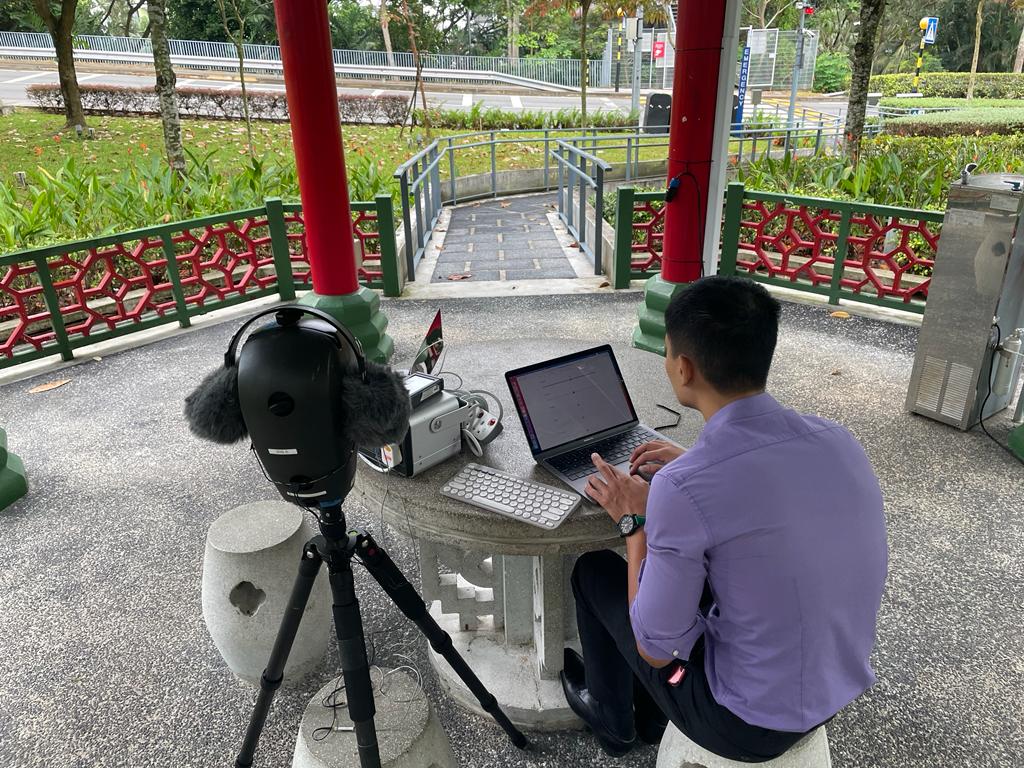}
         \caption{}
         \label{fig:sitenear}
     \end{subfigure}
        \caption{The (a) overview and (b) close up of the binaural measurement system and a participant taking part in the survey in situ.}
        \label{fig:site}
\end{figure}

Fifteen participants, 6 (\SI{40.0}{\percent}) females and 9 males (\SI{60.0}{\percent}), all with normal hearing as judged by the uHear mobile app \cite{Ong2022}, took part in this pilot study. All participants were between 21 and 50 years old, across 3 age bands: 21--30 ($n=9$, \SI{60.0}{\percent}), 31--40 ($n=4$, \SI{26.67}{\percent}), and 41--50 ($n=2$, \SI{13.33}{\percent}); and were recruited within the university campus. The soundscape intervention under test was installed in a Chinese-styled gazebo at Yunnan Garden, Nanyang Technological University, Singapore, as shown in \Cref{fig:site}. The participants were oriented to face the minor road with an unsignalized pedestrian crossing \SI{20}{\meter} from the gazebo. An 8-lane expressway runs parallel to the minor road, about \SI{3}{\meter} below a berm and \SI{45}{m} away from the gazebo. The temperature ($\mu_\text{T}=\SI{31.48}{\celsius}$, $\sigma_\text{T}=\SI{0.84}{\celsius}$) and humidity ($\mu_\text{RH}=\SI{71.94}{\percent}$, $\sigma_\text{RH}=\SI{3.78}{\percent}$) were relatively stable across the entire period of data collection.

Formal ethical approval was sought from the Institutional Review Board (IRB) of NTU (Ref. IRB-2023-399) for this study. In compliance with ethical procedures, informed consent was obtained from all the participants. 

\subsection{Experimental Design}
\noindent
To capture the sound environment as experienced by the participants, a calibrated binaural microphone (TYPE 4101-B, Hottinger Brüel \& Kjær A/S, Virum, Denmark) was mounted on an artificial head (KU 100, Georg Neumann GmbH, Berlin, Germany) with an ear height of \SI{1.2}{\meter}, \SI{0.7}{\meter} from the participant. Due to the outdoor setting, windscreens were used (Windschutz, Soundman e.K., Berlin, Germany), and the binaural acoustic data was recorded with a data acquisition device (SQobold, HEAD acoustics GmbH, Herzogenrath, Germany). 

Participants were instructed to listen to each stimuli (masker) in silence before commencing the evaluation on an electronic form (Qualtrics, Provo, UT, USA). Audio cues were incorporated to indicate the start of the soundtrack and evaluation period, i.e. ``Next soundtrack starting in... 3... 2... 1'' and ``Evaluation starting in... 3... 2... 1...'', respectively. The stimuli would loop continuously in the background during the evaluation period. Once evaluation is completed, the participants were free to advance to the next stimuli by hitting any key on a Bluetooth keyboard. 

The participants first instructed to listen to the surround sound environment by noticing far and near sounds, during which the survey form is frozen. After the end of the 30-s stimuli, participants were prompted to evaluate the surrounding sound environment, while still being exposed to the stimuli, in terms of its perceived affective quality (PAQ) through:

\begin{quote}
    ``\textit{To what extent do you agree or disagree that the present surrounding sound environment is [...]}''
\end{quote}

\noindent
where [...] is one of the eight PAQ attributes (i.e. \textit{eventful}, \textit{vibrant}, \textit{pleasant}, \textit{calm}, \textit{uneventful}, \textit{monotonous}, \textit{annoying}, \textit{chaotic}). The PAQ was judged on a 101-point sliding scale, from \textit{``Strongly disagree''} ($0$) to \textit{``Strongly agree''} ($100$). That is followed by the evaluation of the appropriateness by:

\begin{quote}
   \textit{ ``To what extent is the present surrounding sound environment appropriate to the present place?''}
\end{quote}

\noindent
also on a 101-point sliding scale from \textit{``Not at all''} ($0$) to \textit{``Perfectly''} ($100$).

\subsection{Stimuli}
\noindent
The 6 30-s audio stimuli under test were namely, (1) \ttamb: the in-situ ambient sound environment; (2) \ttbprior: bird masker from \cite{Hong2021b}; (3) \ttwprior: water masker from \cite{Hong2021b}; (4) \ttrand: masker randomly selected from same pool of maskers used in the ARAUS dataset \cite{Ooi2023a}, i.e. \texttt{water\_00037}; (5) \ttbest: masker with the overall highest ISO pleasantness score from 25,440 subjective responses in the ARAUS dataset \cite{Ooi2023a}, i.e. \texttt{bird\_00040}; and (6) \ttamss: the masker as determined by AMSS with the highest ISO pleasantness based on a real-time 30-s snapshot of the in-situ ambient sound environment during the listening experiment. Due to the dynamic nature of the environment and the reactive nature of the AMSS, the 6 30-s stimuli were repeated thrice and presented in random order to each participant, forming a total of 24 stimuli. 

The audio stimuli were presented to each participant seated at the stone table at the centre of the gazebo through a 4-channel audio system, as shown in \Cref{fig:site}. The four loudspeakers (Moukey M20-2, DONNER LLC, FL, USA) were arranged in square at a height of \SI{2.8}{\meter}. A custom internet-of-things (IoT)-based infrastructure was designed to deploy the AMSS \cite{Wong2022}, where the same mono-channel audio file was played from each of the 4 speakers.

 Taking reference from prior work \cite{Hong2020b}, the playback sound pressure level (SPL) for the non-AMSS maskers was determined as \SI{3}{\decibelA} below the 10-min equivalent SPL of a single binaural measurement (45BB-7 Head \& Torso, GRAS Sound and Vibration A/S, Holte, Denmark) at the in-situ location before the study on a typical day, i.e. \SI{66.59}{\decibelA}. Hence, all non-AMSS stimuli were calibrated to \SI{63.59}{\decibelA} in a custom soundproof booth using an automated calibration framework \cite{Ooi2021b}. A speaker with the same make and model of that deployed in situ was placed in the booth \SI{1}{\meter} from the head and torso simulator to perform the calibration. The output levels of each speaker were compensated based on the distance from the speakers to the seating position ($+$\SI{7.60}{\decibelA} at \SI{2.4}{\meter}) and combinatory effect of the 4-speaker setup ($-$\SI{6.02}{\decibelA}) to achieve the desired SPL levels at the listening position.

\section{RESULTS AND DISCUSSION}
\noindent

\subsection{Difference in ISOPL between AMSS predictions and participant scores over time}

\begin{figure}[h!]
     \centering
     \includegraphics[width=0.8\textwidth]{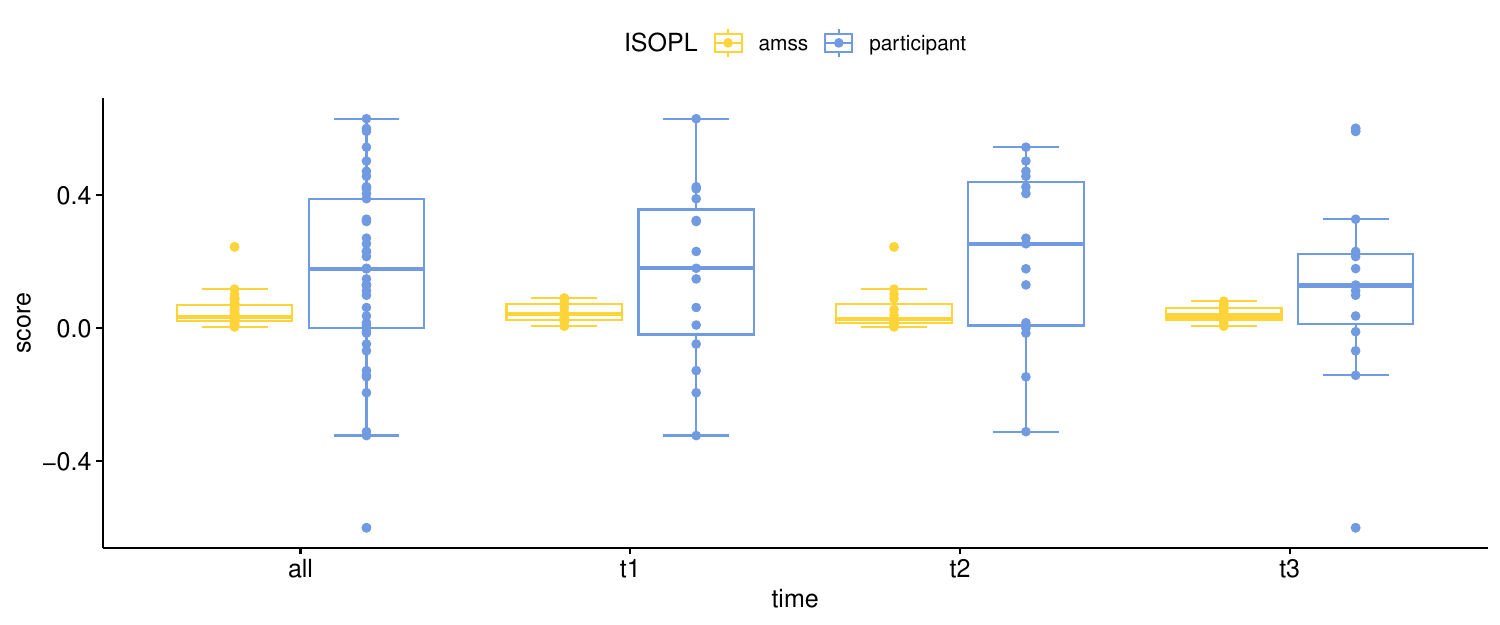} 
    \caption{Boxplot of ISOPL scores across the 3 repetitions between the AMSS predictions and the participant scores.}
        \label{fig:bxpamsspl}
\end{figure}

Both the ISOPL from the AMSS predictions and participant scores (i.e. \ttamss) were summarised in \Cref{fig:bxpamsspl}. To determine if the ISOPL predicted by the AMSS (i.e. soundscape augmented by the masker chosen by AMSS) deviated significantly from the in-situ evaluations by the participants across all 3 time periods, a mixed analysis-of-variance (ANOVA) approach was adopted. 

Two-way mixed ANOVA with time and evaluation method (i.e. AMSS, participant) as independent variables, revealed significant main effects only for evaluation method ($p<0.05$) with an absence of interaction effects. Posthoc analysis using paired t-test showed significant differences in ISOPL between the AMSS predictions and participant scores ($p<0.01$).

The discrepancy in ISOPL between the predictions and in-situ evaluations could be attributed to the mismatch between the microphones in capturing the ambient sound environment in-situ (planar microphone array \cite{Wong2022}) and the training dataset (HATS \cite{Ooi2023a}). Moreover, the results should be interpreted with caution due to the small sample size of this pilot study.

\subsection{Effect of dynamic in-situ sound environment and masker type on soundscape evaluation}
 
\begin{figure}[h!]
     \centering
     \includegraphics[width=\textwidth]{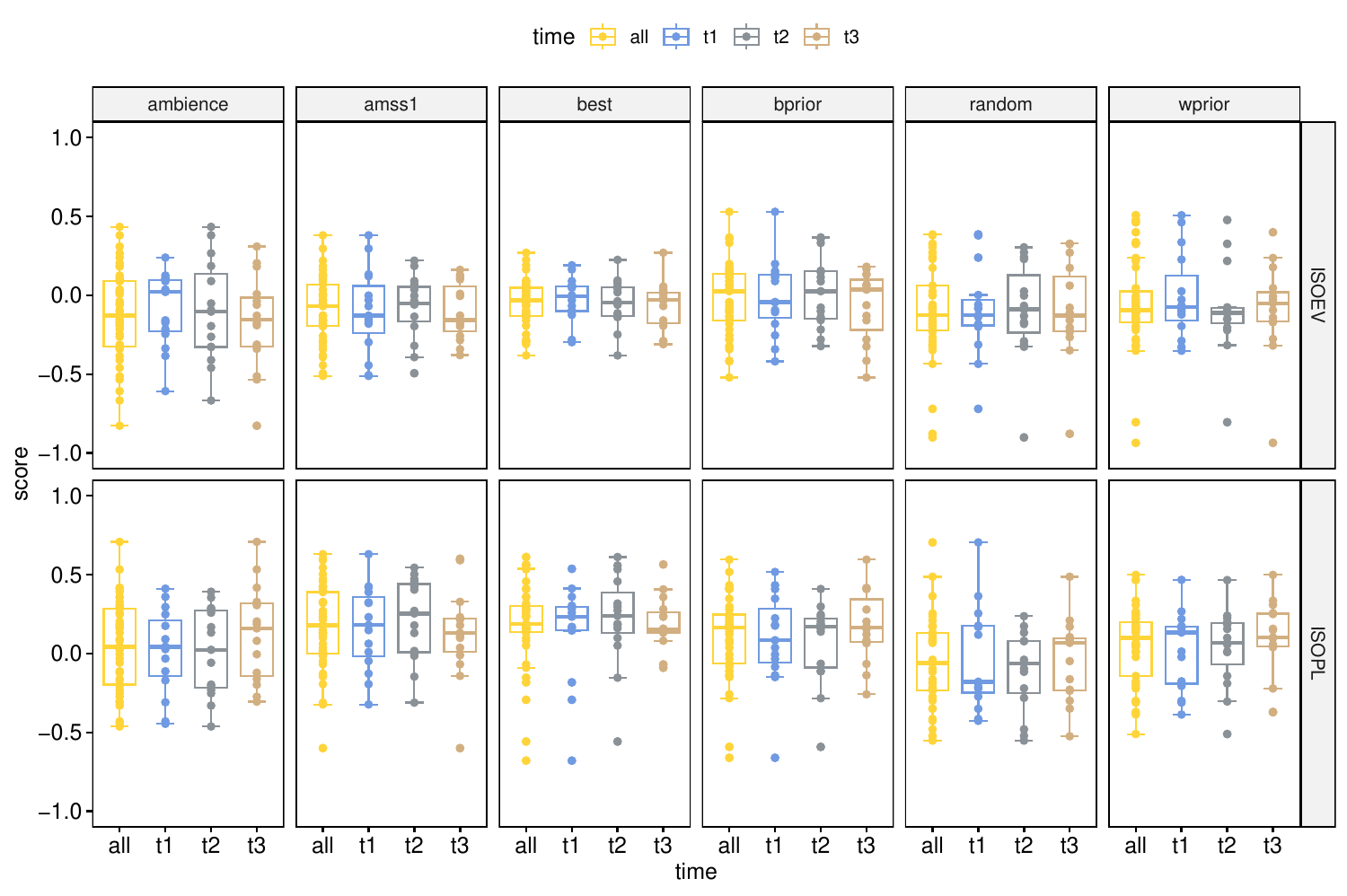} 
    \caption{Boxplot of ISOPL and ISOEV scores across the 3 repetitions for all stimuli types.}
        \label{fig:bxpplev}
\end{figure}

\begin{figure}[h!]
     \centering
     \includegraphics[width=\textwidth]{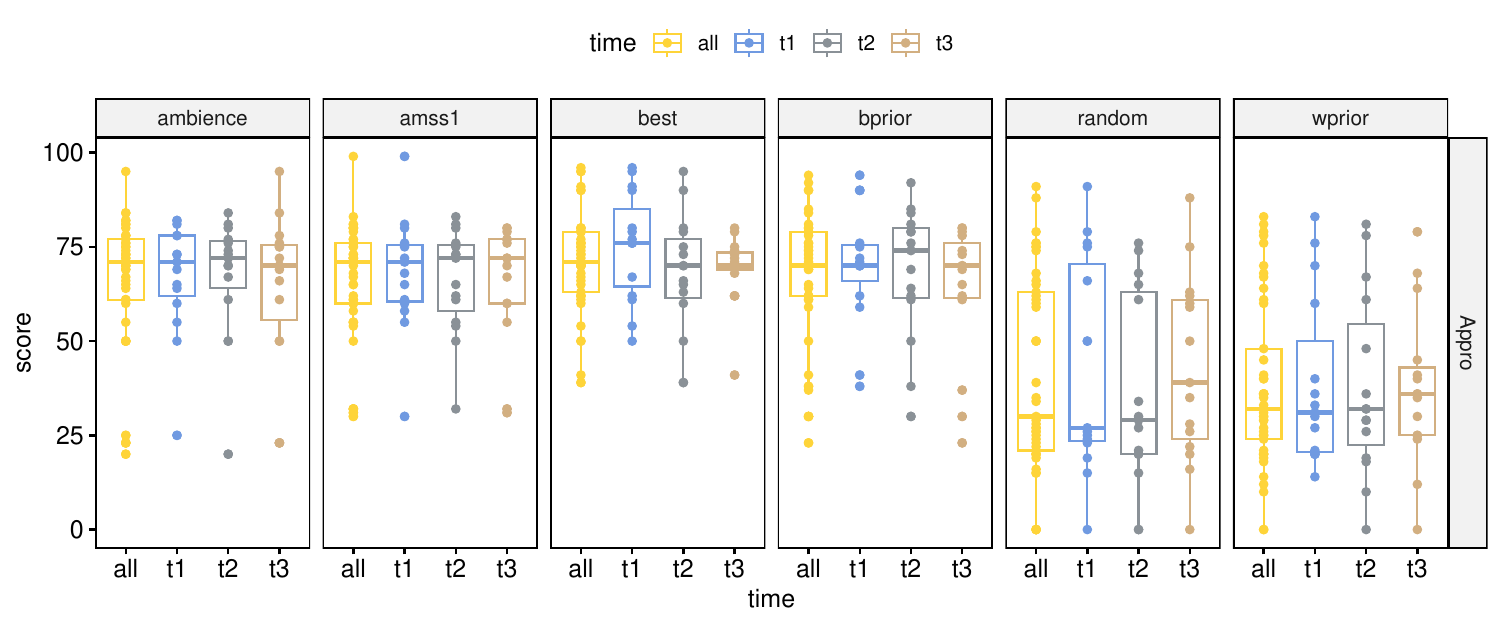} 
    \caption{Boxplot of appropriateness (\textsc{appro}) scores across the 3 repetitions for all stimuli types.}
        \label{fig:bxpappro}
\end{figure}

 The participant scores were summarised by stimuli and time segment for the PAQ (i.e. ISOPL and ISOEV) and appropriateness in \Cref{fig:bxpplev} and \Cref{fig:bxpappro}, respectively. To investigate the effect of the changing sound environment on the soundscape evaluation, two-way repeated measures analysis of variance (2W-RMANOVA) were performed independently across three dependent variables (ISOPL, ISOEV, \textsc{appro}) using time and masker types as independent variables. Main effects were found only for masker types across ISOPL ($p<0.05$) and \textsc{appro} ($p<0.0001$), whereas no interaction effects were found across all dependent variables. 

Posthoc analysis with Tukey's Honest Significant Difference (HSD) test for ISOPL revealed significant difference for pairwise comparisons in \ttrand--\ttamss ($p<0.01$); \ttrand--\ttbest ($p<0.001$); and \ttrand--\ttbprior ($p<0.05$), where the ISOPL with the \ttrand masker was significant lower in all cases.

Posthoc Tukey's HSD test for \textsc{appro} revealed significant difference for pairwise comparisons in \ttrand--\ttamb ($p<0.0001$), \ttwprior--\ttamb ($p<0.0001$), \ttrand--\ttamss ($p<0.0001$), \ttwprior--\ttamss ($p<0.0001$), \ttrand--\ttbest ($p<0.0001$), \ttwprior--\ttbest ($p<0.0001$), \ttrand--\ttbprior ($p<0.0001$), \ttwprior--\ttbprior ($p<0.0001$).

From the 2W-RMANOVA results, there was insufficient evidence to suggest that the dynamic environment had a significant impact on the evaluation of the soundscape across all stimuli. Despite \ttbest ($\mu_\text{ISOPL,best}=0.183$, $\sigma_\text{ISOPL,best}=0.261$), \ttamss ($\mu_\text{ISOPL,amss}=0.166$, $\sigma_\text{ISOPL,amss}=0.268$), and \ttbprior ($\mu_\text{ISOPL,bprior}=0.111$, $\sigma_\text{ISOPL,bprior}=0.261$) exhibiting higher mean ISOPL scores than the \ttamb $\mu_\text{ISOPL,ambience}=0.183$, $\sigma_\text{ISOPL,ambience}=0.261$, none of the augmented soundscapes (i.e. with maskers) reported significantly higher ISOPL over the ambient environment segments, in the pairwise comparisons. 

Unsurprisingly, the \ttrand masker -- \texttt{water\_00037}: rain and thunder -- resulted in a significantly lower ISOPL as compared to \ttamss, \ttbest, and \ttbprior, which could be attributed to its lack of appropriateness (i.e. v.s. \ttamss, \ttbest, \ttbprior, and \ttamb). Similarly, the other water-based masker -- \ttwprior: fountain sound -- was found to be inappropriate in the paired comparison with \ttamss, \ttbest, \ttbprior, and \ttamb. These findings align with the literature, as the absence of water visibility or the sound emitting source led to significantly lower appropriateness \cite{Jeon2012,Hong2020b,Hong2017i,Hong2021b}. 

\section{FINAL COMMENTS AND CONCLUSIONS}

\noindent
 The results of this study provide valuable insights into the short-term in-situ performance of the Automatic Masker Selection System (AMSS) and the effect of various masker selection schemes on perceived affective quality in the dynamic in-situ acoustic environments.

Regarding the difference in ISOPL between the AMSS predictions and participant scores over time, a mixed ANOVA and subsequent posthoc t-tests revealed significant differences in ISOPL between the AMSS predictions and participant scores. The observed discrepancy can be attributed to the mismatch between the microphones used in capturing the ambient sound environment in-situ and those used in the AMSS training dataset.

Furthermore, the effect of the dynamic in-situ sound environment and masker type on soundscape evaluation was examined. The 2W-RMANOVA showed significant main effects for masker types across ISOPL and appropriateness. Posthoc analysis with Tukey's Honest Significant Difference (HSD) test indicated significant differences in ISOPL and appropriateness between various pairwise comparisons. Specifically, the \ttrand masker (i.e. rain and thunder) resulted in significantly lower ISOPL compared to \ttamss, \ttbest, and \ttbprior, which could be attributed to its significant lack of appropriateness.

Overall, the results suggest that the AMSS predictions deviated from in-situ evaluations, and along with the limited sample size, indicates the need for further investigation. Additionally, the choice of masker type significantly influenced the perceived acoustic quality and appropriateness of the soundscape.

However, the findings of this pilot study are limited to the perception of stimuli within a short-term exposure period of 30 seconds. It is important to consider a longer exposure time to investigate the effect of a time-varying masker (e.g. AMSS updates the masker every \SI{30}{\second}) in comparison to the monotony of the same masker being played throughout.  

\section*{Acknowledgements}
\noindent
This work was supported by the National Research Foundation, Singapore, and Ministry of National Development, Singapore under the Cities of Tomorrow R\&D Program (CoT Award: \mbox{COT-V4-2020-1}). Any opinions, findings and conclusions or recommendations expressed in this material are those of the authors and do not reflect the view of National Research Foundation, Singapore, and Ministry of National Development, Singapore.

\bibliographystyle{unsrt}
\bibliography{references}

\end{document}